\documentclass[conference]{IEEEtran}
\IEEEoverridecommandlockouts
\usepackage{cite}
\usepackage{amsmath,amssymb,amsfonts}
\usepackage{algorithmic}
\usepackage{graphicx}
\usepackage{textcomp}
\usepackage{url}
\usepackage{xcolor}
\usepackage{multirow}

\def\BibTeX{{\rm B\kern-.05em{\sc i\kern-.025em b}\kern-.08em
    T\kern-.1667em\lower.7ex\hbox{E}\kern-.125emX}}
\begin{document}

\title{TurnFSM for Full-Duplex Dialogue System: Internalizing State-Machine Logic for Streaming Semantic Voice Activity Detection and Utterance-Level Rejection
}

\author{\IEEEauthorblockN{1\textsuperscript{st} Zhiwei Lin}
\IEEEauthorblockA{\textit{Tsinghua University} \\
Shenzhen, China \\ 
lzw22@mails.tsinghua.edu.cn }
\and
\IEEEauthorblockN{2\textsuperscript{nd} Tianjiao Du}
\IEEEauthorblockA{\textit{Tsinghua University} \\
Shenzhen, China  \\
dtj22@mails.tsinghua.edu.cn
}
\and
\IEEEauthorblockN{3\textsuperscript{rd} Qiaochu Huang}
\IEEEauthorblockA{\textit{Tsinghua University} \\
Shenzhen, China \\
hqc22@mails.tsinghua.edu.cn
}
\and
\IEEEauthorblockN{4\textsuperscript{th} Zihan Zhang }
\IEEEauthorblockA{\textit{Huawei Technologies Co., Ltd.} \\
Shenzhen, China \\
zhangzihan@huawei.com
}
\and
\IEEEauthorblockN{5\textsuperscript{th} Naijun Zheng }
\IEEEauthorblockA{\textit{Huawei Technologies Co., Ltd.} \\
Shenzhen, China \\
njzheng.cuhk@gmail.com \\
}
\and
\IEEEauthorblockN{6\textsuperscript{th} Longshuai Xiao}
\IEEEauthorblockA{\textit{Huawei Technologies Co., Ltd.} \\
Shenzhen, China \\
xiaolongshuai@huawei.com
}
\and
\IEEEauthorblockN{7\textsuperscript{th} Yunfei Lu}
\IEEEauthorblockA{\textit{Huawei Technologies Co., Ltd.} \\
Shenzhen, China \\
luyunfei6@huawei.com
}
\and 
\IEEEauthorblockN{8\textsuperscript{th} Jun Chen}
\IEEEauthorblockA{\textit{Tsinghua University} \\
Shenzhen, China \\
y-chen21@mails.tsinghua.edu.cn
}
\and 
\IEEEauthorblockN{9\textsuperscript{th} Zhiyong Wu}
\IEEEauthorblockA{\textit{Tsinghua University} \\
Shenzhen, China \\
zywu@se.cuhk.edu.hk 
}
}

\maketitle

\begin{abstract}
Full-duplex voice assistants must continuously listen while speaking, handling user interruptions under low-latency and resource-constrained streaming conditions. Existing end-to-end full-duplex models can compromise reasoning-related capabilities after speech-domain adaptation, whereas cascaded pipelines introduce extra inference overhead and handcrafted control logic. We propose TurnFSM, an LLM-based state prediction framework that internalizes turn control as explicit finite-state transitions, unifying streaming semantic VAD and utterance-level rejection. TurnFSM decomposes submission and rejection into a serial decision process, reducing multi-task interference while maintaining performance comparable to single-task models. We further introduce a first-order state transition mechanism that enforces the dependency on only the previous state during training, enabling compact inference with the standard causal mask and original LLM positional encoding while avoiding historical state-token accumulation and unnecessary step-by-step state generation. Experimental results show that TurnFSM consistently outperforms the binary-head baseline and remains competitive with task-specific models.

\end{abstract}


\section{Introduction}
Real-time voice assistants are moving from half-duplex interaction to full-duplex interaction, where the system must continuously listen to user input while generating speech output. This enables natural turn-taking and user interruptions at any time, but also makes input-side control substantially more challenging. In full-duplex scenarios, user speech may overlap with system echoes, background conversations, and environmental noise. Therefore, the system must perform reliable turn control under low-latency and resource-constrained streaming conditions, including speech segmentation, submission timing, and rejection of invalid or unintended inputs.

Existing full-duplex systems generally follow two implementation paradigms. One line of work adopts an end-to-end approach, where streaming turn control and dialogue generation are modeled within a unified speech-language model~\cite{yu2024salmonn,defossez2024moshi,roy2026personaplex,hu2025salm,zeng2024glm}. This paradigm provides a clean interface and can naturally handle overlaps and interjections. However, aligning large language models with the speech modality may affect the textual reasoning ability acquired during pretraining~\cite{wang2026closing, cuervo2025closing, wang2023blsp, zhang2025soundwave}. In addition, semantically equivalent speech and text inputs may still lead to inconsistent internal representations or output distributions, causing cross-modal misalignment. These issues can degrade the stability of reasoning-related capabilities, such as complex reasoning and tool calling, in end-to-end full-duplex models.

In contrast, industrial systems more commonly adopt a cascaded control pipeline~\cite{chen2025minmo, chen2025fireredchat, liu2025x}. An acoustic VAD first segments the continuous audio stream into speech regions. A semantic VAD~\cite{zhang2025llm} then determines whether the current speech segment is semantically complete; if not, the system waits for additional audio and updates the decision until completion or timeout. Once a segment is submitted, an utterance-level rejection module filters invalid or unintended inputs before forwarding accepted utterances to downstream understanding and generation modules. Although such cascaded pipelines are practical and robust, they introduce additional inference overhead, error propagation across stages, and manually designed state machines to coordinate the control logic across modules.

A natural way to reduce cascading overhead is to use a single streaming model for turn control. A straightforward design is to attach multiple prediction heads to a shared streaming backbone, for example one head for semantic completeness and another for utterance-level rejection. However, semantic VAD and rejection are governed by different decision criteria. Semantic VAD focuses on whether the observed audio prefix forms a semantically complete utterance, whereas rejection determines whether the submitted audio is a valid intent-bearing input and therefore depends on both semantic content and acoustic cues, such as noise, echo, and abnormal speech patterns. Directly optimizing these heterogeneous decisions as parallel labels over a shared representation can introduce gradient conflicts and representation interference~\cite{yu2020gradient, navon2022multi, liu2021conflict}, making it difficult to match the performance and stability of dedicated cascaded systems.

To address this problem, we propose TurnFSM, an LLM-based state prediction framework for full-duplex dialogue control. Instead of treating semantic VAD and rejection as independent parallel predictions, TurnFSM internalizes the external decision logic of a cascaded pipeline into explicit finite-state transitions. It models turn control as a serial process over discrete states, such as listening, submission, acceptance, and rejection. This state-transition formulation decomposes full-duplex control into different decision stages: the model first determines semantic completeness during listening, and then performs rejection only after the segment has been submitted. As a result, TurnFSM reduces the need for a shared representation to encode heterogeneous task-specific cues simultaneously, mitigating cross-task interference while preserving a unified streaming controller.
\begin{figure}
    \centering
    \includegraphics[width=0.95\linewidth]{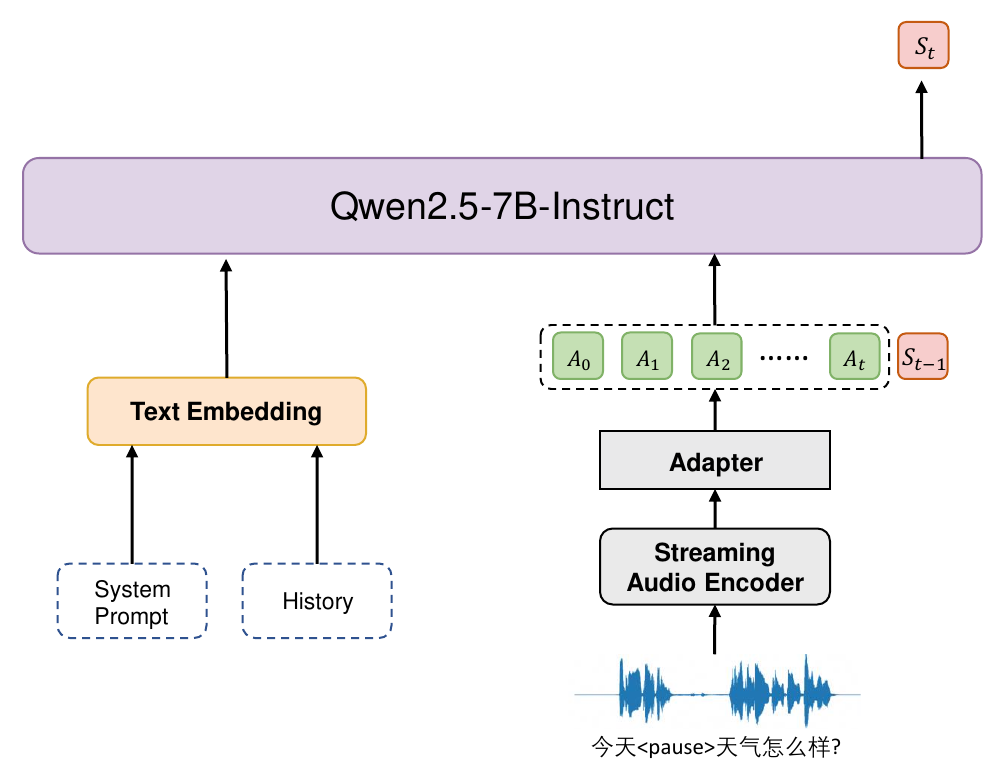}
    \caption{The model architecture of TurnFSM, where $A$ denotes audio tokens and $S$ denotes state tokens. At time step t, the model takes the audio tokens $A_0$, $A_1$, ..., $A_t$ and the predicted state token from the previous time step, $S_{t-1}$, as input to predict the current state token $S_t$.}
    \label{fig:state}
\end{figure}

Furthermore, we introduce a first-order state transition mechanism to enable efficient state prediction. A naive token-level implementation would condition each state prediction on the full history of previous state tokens, i.e., $p_\theta(S_t \mid A_{\le t}, S_{<t})$, requiring historical state tokens to be repeatedly accumulated in the context. In contrast, TurnFSM enforces a first-order transition dependency, $p_\theta(S_t \mid A_{\le t}, S_{t-1})$, where the current state depends only on the acoustic prefix and the immediately preceding state. During training, this dependency is implemented with a modified attention mask and a position-tying strategy over interleaved audio-state sequences. At inference time, the model can use the compact input form $(A_0,\ldots,A_t,S_{t-1})$ and predict $S_t$ with the standard causal attention mask and the original LLM positional encoding scheme. This avoids historical state-token accumulation, removes unnecessary step-by-step state generation, and supports efficient inference in both streaming and utterance-level settings.

In summary, our contributions are as follows:
\begin{itemize}
\item We propose TurnFSM, a streaming full-duplex state predictor that jointly supports semantic VAD and utterance-level rejection within a unified state-transition framework.

\item TurnFSM internalizes the decision logic of cascaded control pipelines through explicit finite-state transitions, decomposing semantic VAD and rejection into different decision stages and reducing cross-task interference caused by parallel multi-task heads.

\item We propose a first-order state transition mechanism that reduces state dependency from the full history $S_{<t}$ to the immediately preceding state $S_{t-1}$, avoiding historical state-token accumulation and unnecessary step-by-step state generation.

\end{itemize}

\section{Methodology}
TurnFSM replaces the handcrafted control state machine in cascaded full-duplex pipelines with a learnable state transition model embedded in the LLM. Given streaming audio, it predicts the current control state conditioned on the previous state, enabling online turn control within a unified state space. TurnFSM further adopts a serial decomposition—submission first, then rejection detection to assign semantic completeness and rejection to different state stages, avoiding the representation competition and gradient interference caused by shared multi-label representations.

\subsection{Model Architecture}

TurnFSM consists of three modules: a streaming audio encoder, an audio adapter, and an LLM backbone. The streaming audio encoder consists of a two-layer VGG-style~\cite{simonyan2014very} convolutional subsampling front-end with 4$\times$ downsampling, followed by a 6-layer Conformer encoder, which encodes continuous audio into a sequence of acoustic representations. The encoder outputs acoustic frames at a rate of 25 Hz, corresponding to one frame every 40 ms. The audio adapter is a lightweight MLP with 2$\times$ downsampling, which projects the acoustic representations into the input embedding space of the LLM. After this additional downsampling, the final audio token rate becomes 12.5 Hz, corresponding to one audio token every 80 ms. We use Qwen2.5-7B-Instruct~\cite{qwen2025qwen25technicalreport} as the LLM backbone for streaming state prediction.

To achieve stable cross-modal alignment and streaming state prediction, we adopt a two-stage training procedure. We initialize the audio encoder from an internally pre-trained streaming automatic speech recognition (ASR) model and treat it as the acoustic backbone; details of its pre-training are omitted as they are not the focus of this work. In the first stage, we freeze both the audio encoder and the LLM and train only the audio adapter for modality alignment. In the second stage, we unfreeze the audio adapter and the LLM and optimize the state prediction objective to enable reliable state transitions under streaming input.
\begin{figure}
    \centering
    \includegraphics[width=0.78\linewidth]{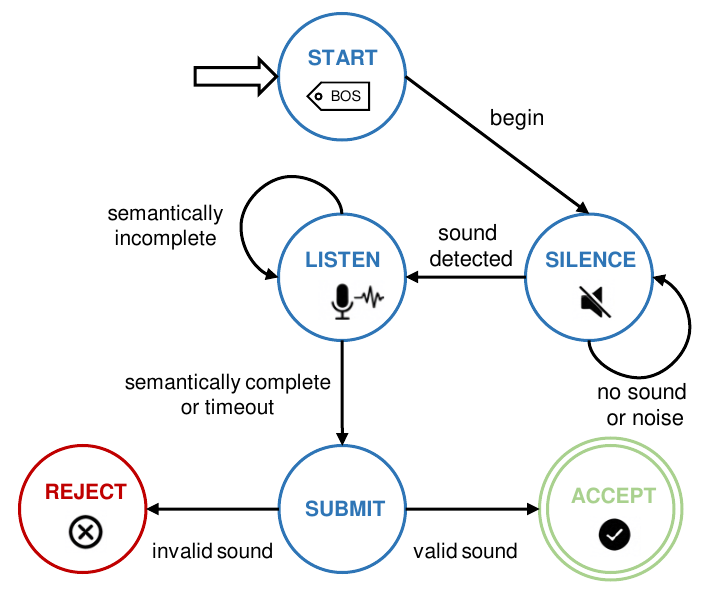}
    \caption{The state transition logic of TurnFSM.}
    \label{fig:state}
\end{figure}
\begin{figure*}[t]
    \centering
    \includegraphics[width=0.95\textwidth]{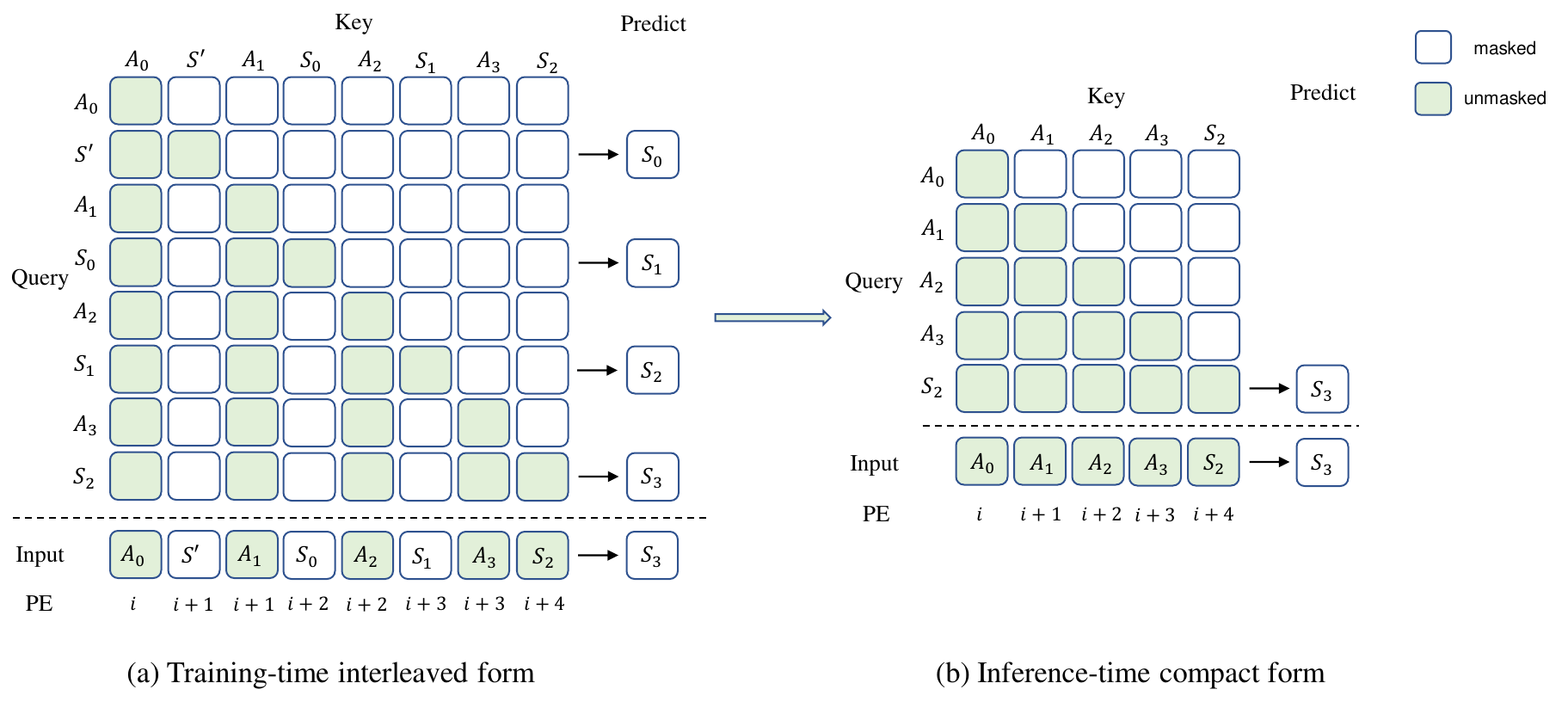}
    \caption{
    Illustration of the proposed masking and position-tying strategy, where $S'$ denotes the pseudo-initial state $<|Start|>$ and PE denotes positional encoding.
    (a) During training, a modified causal mask over the interleaved audio-state sequence restricts each $S_t$ to attend only to $A_{\le t}$ and $S_{t-1}$.
    State tokens are position-tied to the following audio tokens, i.e., $\pi(S_{t-1})=\pi(A_{t+1})$, to preserve the original acoustic temporal order.
    (b) During inference, TurnFSM predicts $S_t$ from the compact sequence $(A_0,\ldots,A_t,S_{t-1})$ using the standard lower-triangular causal mask and the original LLM positional encoding.}
    \label{fig:mask}
\end{figure*}

\subsection{Finite-State Turn Control Formulation}

In streaming full-duplex dialogue control, a straightforward baseline is to attach two binary prediction heads to a shared backbone: one for semantic completeness, i.e., submit versus wait, and the other for input validity, i.e., accept versus reject. This can be formulated as
\[
p_\theta(c_t \mid A_{\le t}), \qquad
p_\theta(r_t \mid A_{\le t}),
\]

where $c_t$ indicates whether the current audio prefix is semantically complete, and $r_t$ indicates whether the accumulated audio should be accepted or rejected.
However, these two decisions correspond to distinct tasks with different decision criteria. The completeness decision is semantic VAD, which determines whether the audio observed up to the current time step forms a semantically complete utterance. In contrast, rejection determines whether the accumulated audio is a valid intent-bearing input. This decision requires not only semantic information, but also acoustic cues, such as noise, echo, abnormal speech patterns, and other non-intent signals. Therefore, optimizing semantic VAD and rejection as independent labels under a shared objective may introduce conflicting gradients in the shared representation space, reducing the stability of streaming control.

As shown in Figure~\ref{fig:state}, TurnFSM explicitly models the control process of the cascaded pipeline as a finite state machine. At each time step $t$, the model maintains a discrete control state $S_t \in \mathcal{S}$, where
\begin{multline}
\mathcal{S} = \{<|Start|>,\, <|Silence|>,\, <|Listen|>,\\
<|Submit|>,\, <|Accept|>,\, <|Reject|> \}.
\end{multline}

The state tokens are defined as follows:
\begin{itemize}
\item $<|Start|>$: A pseudo initial token before any input is processed.
\item $<|Silence|>$: The non-user-speech state, including silence and background noise. In this state, the system continuously monitors the input without buffering any segment.
\item $<|Listen|>$: The active listening state entered after a speech onset is detected. Acoustic frames are accumulated as a candidate user utterance.
\item $<|Submit|>$: The commitment state reached when the buffered segment is predicted to be semantically complete or when a timeout is triggered. The segment is finalized and passed to utterance-level validation.
\item $<|Accept|>$: A terminal state indicating that the submitted audio is a valid user intent and should be handled by the dialogue model.
\item $<|Reject|>$: A terminal state indicating that the submitted audio should be filtered out. This includes both acoustically invalid inputs, such as noise, echo, or electronic sounds, and non-intentional but semantically meaningful audio, such as background conversations or unintended speech.
\end{itemize}

Unlike the dual-head classification scheme, TurnFSM formulates turn control as a finite-state transition process. Conceptually, each transition is defined to depend on the acoustic prefix and the immediately preceding state:
\[
p_\theta(S_t \mid A_{\le t}, S_{t-1}).
\]
This first-order transition formulation decomposes turn control into a sequential decision process, rather than requiring the model to optimize semantic VAD and rejection simultaneously at every time step. Specifically, when $S_{t-1}=<|Listen|>$, the model focuses on semantic completeness and determines whether to transition to $<|Submit|>$. Once $S_{t-1}=<|Submit|>$, the model performs rejection by assessing whether the submitted segment is a valid intent-bearing input, and then transitions to either $<|Accept|>$ or $<|Reject|>$. By separating these decisions across different states, TurnFSM reduces the need for a shared representation to simultaneously encode heterogeneous task-specific cues, thereby mitigating performance degradation caused by conflicting optimization signals.

\subsection{First-Order State Transition Mechanism}

The transition dependency above describes the desired FSM formulation. However, a naive token-level implementation may fail to preserve this dependency during training. If audio tokens and state tokens are simply interleaved and trained with a standard causal attention mask, the prediction of $S_t$ can attend to the entire history of previous state tokens:
\[
p_\theta(S_t \mid A_{\le t}, S_{<t}).
\]
This history-dependent formulation is inefficient and unnecessary for TurnFSM. Earlier state tokens provide limited additional evidence for the current decision, because the immediately preceding state $S_{t-1}$ already specifies the current control phase and determines which type of decision should be made: semantic completeness in the $<|Listen|>$ state, or utterance-level validity after entering the $<|Submit|>$ state. Therefore, feeding back earlier states $S_{<t-1}$ brings little direct benefit to the current decision, while increasing the context length and inference overhead.

This also creates a train--inference consistency issue. If the model relies on interleaved historical state tokens during training, inference would typically need to follow the same chunking and state-update pattern as used during training.
As a result, even for non-streaming or utterance-level inference, the model would still need to generate intermediate state tokens step by step before producing the final semantic completeness or rejection decision. This sequential state-token generation is computationally expensive and unnecessary when only the current state decision is needed.

To enforce the intended first-order dependency while supporting efficient inference, we introduce a first-order state transition mechanism. During training, we construct an interleaved sequence of audio tokens and state tokens, as illustrated in Fig.~\ref{fig:mask}(a). For state-token prediction, we modify the training attention mask such that $S_t$ is allowed to attend only to the acoustic prefix $A_{\le t}$ and the immediately preceding state token $S_{t-1}$, while attention to earlier state tokens $S_{<t-1}$ is masked. This ensures that the token-level training objective matches the desired FSM dependency:
\[
p_\theta(S_t \mid A_{\le t}, S_{t-1}).
\]

We further apply position tying to align the interleaved training sequence with the compact inference form. If positional indices were assigned sequentially to the interleaved sequence, the inserted state tokens would shift the original acoustic temporal order. Therefore, for the training sequence $(A_0,<|Start|>,A_1,S_0,A_2,S_1,\dots)$, we tie each state token to the positional encoding of the following audio token:
\[
\pi(<|Start|>)=\pi(A_1),\qquad \pi(S_{t-1})=\pi(A_{t+1}).
\]
Here, $A_{t+1}$ is used only as a positional reference in the interleaved training sequence, rather than as an input for predicting $S_t$. With this design, audio tokens retain their original positional indices, and each state token is placed at the position it would occupy in the compact inference sequence.

As shown in Fig.~\ref{fig:mask}(b), the trained model can therefore predict $S_t$ from the compact sequence $(A_0,\ldots,A_t,S_{t-1})$ at inference time. Since this sequence follows the standard autoregressive order, TurnFSM directly uses the standard lower-triangular causal attention mask and the original LLM positional encoding scheme. Consequently, the model avoids historical state-token accumulation, removes unnecessary step-by-step state generation, and supports efficient state prediction in both streaming and utterance-level settings.

\section{Experiments}
\begin{table*}[!htbp]
\centering
\caption{Comparison of different models on the streaming semantic VAD and rejection-detection tasks, where SR, TOR, and ECR represent success rate, timeout rate, and early-cut rate, respectively. }
\scalebox{0.9}
{
\begin{tabular}{cccccccc}
\hline
Model & SR(\%) $\uparrow$ & ECR(\%) $\downarrow$ & TOR(\%) $\downarrow$ & Mean Lat. (Succ.)(ms) $\downarrow$ &  Mean Lat. (All)(ms) $\downarrow$ & FAR(\%) $\downarrow$ & FRR(\%) $\downarrow$ \\
\hline
TurnFSM & \textbf{82.41} & \textbf{12.06} & \textbf{5.53} & 80.9 & \textbf{222.7} & 3.35 & \textbf{16.04} \\
Binary-Head & 80.29 & 12.06 & 7.67 & 119 & 282.9 & \textbf{3.27} & 18.38 \\
TEN Turn Detection & 58.75 & 30.25 & 11 & 117.57 & 414.43 & - & - \\
Smart Turn V3.2 & 68 & 26.75 & 5.25 & 643.43 & 693.11 & - & - \\
Easy Turn & 69.5 & 22.25 & 8.25 & 79.12 & 270.08 & - & - \\
VAD-only & 81.9 & 11.55 & 6.55 & \textbf{77.83} & 232.6 & - & - \\
Rejection-only & - & - & - & - & - & 3.31 & 16.16\\
\hline
\end{tabular}
}

\label{tab:lab1}
\end{table*}
\subsection{Experimental Setup}
TurnFSM is trained in two stages: cross-modal alignment and state prediction.
For cross-modal alignment, we train a bridging layer using open-source speech corpora on an ASR task, including AISHELL-1\cite{bu2017aishell}, AISHELL-2\cite{du2018aishell}, and WenetSpeech\cite{zhang2022wenetspeech}, totaling 11,200 hours of audio.
For state prediction, we use two internal datasets. To train the state transition from $<|Start|>$ to $<|Submit|>$, we used an internal semantic VAD dataset consisting of 70K training samples and 800 test samples. The training set totals 194 hours and is evenly split between samples with pauses and without pauses. Each sample is a semantically complete sentence, annotated by humans with acoustic start and acoustic end timestamps.
To train the state transition from $<|Submit|>$ to $<|Reject|> / <|Accept|>$, we used an internal rejection-detection dataset containing 700K training samples and 3K test samples. The training set totals 2,000 hours, with an even positive/negative split, and covers a wide range of real-world scenarios. 
The model alignment training phase was run on 16 Ascend 910b NPUs using a learning rate of 6e-5 and a batch size of 256 for two training epochs. For the state prediction phase, the training phase was run on 64 Ascend 910b NPUs using a learning rate of 2e-5 and a batch size of 64 for four training epochs.
\subsection{Metircs and Baselines}
To effectively evaluate TurnFSM on the semantic VAD task, we report the following metrics:
\begin{itemize}

\item Early-cut rate: the proportion of utterances where the predicted semantic end point occurs earlier than the human-labeled end point. Since manual endpoint annotation is inherently noisy, we apply a 100 ms tolerance window: predictions within ±100 ms of the labeled endpoint are treated as correct and are not counted as early cuts.

\item Timeout rate: the proportion of semantically complete utterances that are incorrectly judged as incomplete by the model, leading to a timeout.

\item Success rate: the proportion of utterances excluding early-cut and timeout cases.

\item Mean latency on successful cases: the average time difference between the predicted semantic endpoint and the labeled endpoint among successful predictions.

\item Overall mean latency: the average latency over both successful and timeout cases, where timeout cases are assigned a fixed latency of 2000 ms.
\end{itemize}
In addition, for rejection detection, we use false rejection rate (FRR) and false acceptance rate (FAR) as evaluation metrics.

To evaluate the effectiveness of state-transition modeling, we compare TurnFSM against an alternative approach that uses a binary classification head.
In addition,  we compare TurnFSM with three open-source semantic VAD models, including TEN Turn Detection\cite{TEN_Turn_Detection}, Easy Turn \cite{li2025easy} and Smart Turn V3.2\footnote{\url{https://github.com/pipecat-ai/smart-turn}}, to assess its semantic VAD performance.
Since TEN Turn Detection is a text-based prediction model, we use Paraformer\cite{gao2022paraformer} as the ASR model to transcribe speech into text and feed the transcripts as its input.
Notably, all compared open-source baselines are offline models and do not directly produce online decision timestamps. To enable latency evaluation under a streaming setting, we simulate streaming inference by first segmenting the audio into speech chunks using Silero VAD~\cite{SileroVAD}, and then feeding these chunks to each model incrementally.

TEN Turn Detection and Easy Turn output only a binary decision, which naturally supports a fixed control policy, i.e., either submitting the current chunk or continuing to wait. Such binary outputs make it difficult to implement fine-grained adaptive waiting. In contrast, Smart Turn produces a continuous semantic completeness probability $p$, which indicates how likely the current audio is to be semantically complete. Therefore, for Smart Turn, we adopt a probability-driven adaptive waiting strategy. Specifically, for each predicted probability $p$, we convert it into a waiting time:
\[
t_{\mathrm{wait}} = (1 - p) \times 2000~\mathrm{ms}.
\]
If a new speech chunk arrives within $t_{\mathrm{wait}}$, we merge it with the current chunk and re-evaluate the merged audio. Otherwise, we treat the current chunk as semantically complete and trigger submission.

\subsection{Main Result}

As shown in Table~\ref{tab:lab1}, TurnFSM achieves performance comparable to the corresponding single-task models on both semantic VAD and rejection detection. On semantic VAD, TurnFSM obtains a higher success rate than the VAD-only model (82.41\% vs. 81.90\%) and a lower timeout rate (5.53\% vs. 6.55\%), while maintaining a similar early-cut rate and latency. On rejection detection, TurnFSM also performs close to the Rejection-only model, achieving a comparable FAR (3.35\% vs. 3.31\%) and a slightly lower FRR (16.04\% vs. 16.16\%). These results indicate that the state-machine formulation effectively decouples semantic VAD and rejection into different transition stages, allowing the model to focus on task-relevant information without substantially sacrificing either subtask.

In contrast, the binary-head baseline performs worse than TurnFSM on most metrics. Although it uses the same shared backbone, directly optimizing semantic VAD and rejection as two independent prediction heads leads to a lower success rate, a higher timeout rate, larger latency, and a higher FRR. This suggests that the dual-head formulation introduces stronger multi-task coupling in the shared representation space, which may weaken the optimization of each subtask and degrade overall streaming control performance.

Compared with open-source semantic VAD baselines, TurnFSM achieves the best overall balance. It obtains the highest success rate (82.41\%) while maintaining competitive early-cut and timeout rates. Although Easy Turn achieves a slightly lower Mean Lat. (Succ.), its higher timeout rate leads to a larger Mean Lat. (All) when timeout cases are included. Smart Turn achieves a timeout rate comparable to TurnFSM, mainly because its probability-driven adaptive waiting strategy forces a decision once the waiting time is exceeded. However, this strategy is more conservative in successful cases, resulting in substantially higher Mean Lat. (Succ.) and Mean Lat. (All). Overall, TurnFSM provides more stable submission decisions with lower overall latency, while also supporting rejection detection within the same state-transition framework.

\begin{table}[!htbp]
\centering
\caption{Comparison of utterance-level semantic VAD performance on the EasyTurn test set, which contains semantically complete and incomplete utterance subsets. $\mathrm{ACC}_{\mathrm{cp}}$, $\mathrm{ACC}_{\mathrm{incp}}$, and $\mathrm{ACC}_{\mathrm{all}}$ denote the accuracy on complete samples, incomplete samples, and all samples, respectively.}
\scalebox{0.85}
{
    \begin{tabular}{cccc}
    \hline
    Model & $ACC_{cp}(\%) \uparrow$ & $ACC_{incp}(\%) \uparrow$ & $ACC_{all}(\%) \uparrow$ \\
    \hline
    TurnFSM & 92 & 90 & 91\\
    TEN Turn Detection & 86.67 &  89.3 & 87.99\\
    Smart Turn V3.2 & 80 & 69.33 & 74.665 \\
    Easy Turn & \textbf{96.33} & \textbf{97.67} & \textbf{97} \\
    \hline
    \end{tabular}
}
\label{tab:lab2}
\end{table}

In addition, we evaluate utterance-level semantic VAD performance on the open-source EasyTurn test set, which consists of two subsets: semantically complete utterances and incomplete utterances. Different from streaming evaluation, this benchmark does not require online decision timestamps; instead, the model only needs to determine whether each input utterance is semantically complete.

This setting also highlights the efficiency advantage of the proposed first-order state transition mechanism. Although TurnFSM is trained with interleaved audio-state sequences, the masking design makes the learned transition compatible with the compact inference form $(A_0,\ldots,A_t,S_{t-1})$. Therefore, for utterance-level evaluation, TurnFSM can directly process the whole input utterance and predict the final state in a single forward pass, without step-by-step generation of historical state tokens.

As shown in Table~\ref{tab:lab2}, TurnFSM achieves an overall accuracy of 91.00\%, outperforming TEN Turn Detection (87.99\%) and Smart Turn V3.2 (74.67\%). Although it still lags behind Easy Turn (97.00\%), TurnFSM obtains the second-best result on this benchmark. These results indicate that TurnFSM generalizes well to utterance-level semantic completeness prediction, while retaining efficient inference enabled by the proposed masking mechanism.

\subsection{Ablation Study}
\begin{table}[!htbp]
\centering
\caption{Performance comparison of ablation components on semantic VAD task, where FOSTM denotes the proposed first-order state transition mechanism.}
\scalebox{0.82}{
\begin{tabular}{cccccc}
\hline
Model & SR(\%) & ECR(\%) & TOR(\%) & Mean Lat.(Succ.)(ms) \\
\hline
TurnFSM & \textbf{82.41} & \textbf{12.06} & 5.53 & 80.9 \\
w/o FOSTM & 81.97 & 13.32 & \textbf{4.71} & 111.35 \\
w/o cross-modal & \multirow{2}{*}{79.39} & \multirow{2}{*}{15.83} & \multirow{2}{*}{4.78} & \multirow{2}{*}{\textbf{78.3}} \\
alignment       &                        &                        &                       &                                   \\
\hline
\end{tabular}
}
\label{tab:lab3}
\end{table}
Tables~\ref{tab:lab3} and~\ref{tab:lab4} present the ablation results of the proposed components. We first analyze the effect of the first-order state transition mechanism (FOSTM). Removing FOSTM leads to a slightly lower success rate on semantic VAD and a higher early-cut rate, while also increasing the successful-case latency from 80.9 ms to 111.35 ms. On the rejection-detection task, the performance gap is small, with only minor changes in FAR and FRR. These results show that FOSTM preserves the modeling capability of the original FSM formulation while maintaining stable performance across both tasks.
\begin{table}[!htbp]
\centering
\caption{Performance comparison of ablation components on rejection-detection task, where FOSTM denotes the proposed first-order state transition mechanism. }
\scalebox{0.9}{
\begin{tabular}{cccccc}
\hline
Model & FAR(\%) & FRR(\%) \\
\hline
TurnFSM & 3.35 & \textbf{16.04}\\
w/o FOSTM & 3.46 & 16.34  \\
w/o cross-modal & \multirow{2}{*}{\textbf{3.27}} & \multirow{2}{*}{17.10} \\
alignment       &                        &              \\

\hline
\end{tabular}
}
\label{tab:lab4}
\end{table}

More importantly, FOSTM provides a clear inference-efficiency advantage. Without FOSTM, the model has to follow the naive interleaved state-token formulation, where historical state tokens are repeatedly fed back into the context. This not only increases the input length over time, but also forces inference to follow the fixed training-time state-update pattern. In contrast, FOSTM enforces the first-order dependency $p_\theta(S_t \mid A_{\le t}, S_{t-1})$ during training, allowing TurnFSM to use the compact inference form $(A_0,\ldots,A_t,S_{t-1}) \rightarrow S_t$ with the standard causal mask. Therefore, FOSTM reduces redundant state-token accumulation, avoids unnecessary step-by-step state generation, and enables efficient inference in both streaming and utterance-level settings.

We also evaluate the effect of cross-modal alignment. Removing the alignment stage causes a noticeable degradation on semantic VAD, especially increasing the early-cut rate from 12.06\% to 15.83\%. This suggests that alignment helps the model learn audio representations that are more consistent with semantic content, which is crucial for detecting semantic completeness and avoiding premature submission. In contrast, its impact on rejection detection is relatively limited: the FAR slightly decreases, but the FRR increases from 16.04\% to 17.10\%. This is reasonable because rejection detection depends not only on semantic content, but also on acoustic cues such as noise, echo, and abnormal speech patterns. Therefore, semantic-oriented cross-modal alignment benefits semantic VAD more directly than rejection detection.

\section{Conclusion}
In this paper, we proposed TurnFSM, a streaming state prediction framework for full-duplex voice assistants. TurnFSM unifies semantic VAD and utterance-level rejection through explicit finite-state transitions, transforming parallel multi-task prediction into a serial decision process. By separating submission and rejection into different states, TurnFSM reduces cross-task interference and allows the model to focus on task-relevant cues at each transition. We further introduced a first-order state transition mechanism, which aligns interleaved training with compact inference and avoids historical state-token accumulation during deployment. Experimental results show that TurnFSM achieves strong semantic VAD and rejection performance, performs competitively with task-specific models, and improves over the binary-head baseline. These results demonstrate that TurnFSM is an effective and practical low-latency plug-in module for full-duplex dialogue systems.




\clearpage
\bibliographystyle{IEEEtran}
\bibliography{mybib}

\end{document}